\documentstyle[11pt,pasp,twoside]{article}
\def\edcomment#1{\iffalse\marginpar{\raggedright\sl#1\/}\else\relax\fi}
\reversemarginpar

\def\phfl{phot$\cdot$cm$^{-2}$s$^{-1}$ }
\def\*{$^{*}$}

\begin{document}
\title{Non-pulsing emission from X-ray pulsars}
\author{Lutovinov A., Grebenev S., Sunyaev R.}
\affil{Space Research Institute, Profsoyuznaya 84/32, 117810 Moscow,
  Russia} 

\begin{abstract}
Results of GRANAT/ART-P observations of three X-ray pulsars in non-pulsing
states are presented: 1) a statistically significant non-pulsing flux with a
simple power-law spectrum was detected during the ``off''- state of Her X-1;
2) a significant (13 $\sigma$) non-pulsing flux with a strong iron emission
line at energies ~ 6.7-6.9 keV was detected during the eclipse ingress of
Cen X-3; 3) a weak non-pulsing flux was detected during the X-ray eclipse of
Vela X-1, which probably resulted from scattering of the pulsar emission in
the stellar wind of an optical star.
\end{abstract}

\section{Her X-1}

The ``off'' state of the X-ray pulsar Her X-1 was observed with the
telescope ART-P on board GRANAT three times in 1990-1991. The source
persistent flux in this state was detected at the level of few mCrabs
(several percents of its high-state flux) and was characterized by the
absence of X-ray pulsations --- the $3\sigma$ upper limit on the pulse
fraction in the 3-20 keV band varied between 2.9\% and 18\% in different
observations. The source's ``off'' state spectra are well described by a
simple power law without evidence of a cutoff in the hard energy band or of
emission lines. The spectral slopes differ by a factor of $\sim2$, with the
photon flux in the energy band 3-20 keV being virtually the same. Her~X-1 is
observed in its ``off'' state if the compact source is hidden behind the
outer edge of a warped (or tilted to the orbital plane) accretion disk and,
as it is believed, the source emission can greatly scattered in a hot corona
(it's height is $H\ge 10^{11}$ cm) above the outer parts of the disk. Such
scattering of emission from the compact source by coronal electrons can
account for the existence of an X-ray flux in the ``off'' state and it
allows to explain the absence of pulsations in this state, as any
information about variability on a time scale shorter than $H/c\sim 3$ s,
where $c$ is the speed of light, must be lost.

\section{Cen X-3}

A significant declination (by a factor of 6) in the 4-20 keV source flux was
detected during the observation on Aug 19, 1990. This declination occurred in
two steps, latter of which corresponded to an orbital phase $\phi\simeq0.89$
at which the X-ray eclipse ingress was observed in other experiments. It is
important to note that the intensity decline during the eclipse ingress
mainly occurred in the hard energy band and the corresponding hardness
(ratio of the 10-20 keV and 4-6 keV fluxes) decrease was equal to
$\sim40$\%. After the eclipse ingress a non-zero and non-pulsing X-ray
emission was detected by ART-P from the source at the signal-to-noise level
of $S/N\simeq13\sigma$. The pulsar's spectrum during the eclipse was
measured with large statistical errors; nevertheless, a line at energy of
$\sim7$ keV was very reliable. By assuming that this is the 6.7-keV line of
helium-like iron, we estimated its intensity, $I_{\rm 6.7}\simeq
(2.7\pm1.4)\times10^{-3}$\phfl, and equivalent width, $EW_{\rm
6.7}\simeq(1.2\pm0.6)$ keV. Although the source itself was already occulted
by the disk of the optical star at this time, part of the surface of the
scattering cloud remained in the visibility zone, and we observed precisely
this emission scattered in the cloud. Note also that the measured spectrum
at energies below 5 keV shows an excess of soft X-ray emission, which was
previously revealed from the light curve analysis.

\section{Vela X-1}

The observation of this X-ray pulsar on June 15, 1992 (orbital phases
0.918--0.953) was carried out mainly during an eclipse. We divided it into
two parts, in accordance with the beginning of the source's eclipse
egress. A non-zero flux of approximately 1/10 of the bright-state flux at
the confidence levels 4.3 and 4.6$\sigma$ was observed from the source
during the eclipse and eclipse egress, respectively. During the second part
of the observation X-ray pulsations with the period $P=283.33\pm1.65$ s were
detected from the source at a confidence level of $10.2\sigma$. The pulse
fraction was equal to $33.9\pm6.4$\% and it was a factor of 2 lower than one
measured in the source's bright state several days earlier. During the
eclipse we failed to detect any pulsations --- the 3$\sigma$ limit on the
pulse fraction was 9.6\%. The persistent flux, detected during an X-ray
eclipse, may be the emission scattered in a fairly dense stellar wind or in
the extended atmosphere of the optical star. Given that the binary
characteristic size, $l\sim a/c\simeq 120$ light seconds, is half the
pulsation period, photons must undergo several scatterings for the pulse
profile to be smeared enough.  For the scattering optical depth of the wind
$\tau_{\rm T}\sim 1$, the scattered-to-incident flux ratio allows us to
estimate the solid angle $\Omega\sim0.4\pi a^2$ at which the
scattering-envelope ring is seen from the compact source. Accordingly, the
envelope height above the optical-star surface is $H\simeq 14 R_{\odot}.$

More comprehensive discussion of the results of the observations non-pulsing
emission from X-ray pulsars with the ART-P telescope are presented elsewhere
(see Lutovinov et al. 1999, 2000a,b).

\end{document}